\documentclass[11pt,a4]{scrartcl}

\usepackage{cite}
\usepackage[pdftex]{graphicx}

\usepackage{upgreek}
\usepackage{url}
\usepackage{booktabs}

\begin{document}
%

\title{Design and First Tests of a Radiation-Hard Pixel Sensor for the European X-Ray Free-Electron Laser}

\author{Joern~Schwandt \and 
        Eckhart~Fretwurst \and 
        Robert~Klanner \and 
        Ioannis~Kopsalis \and
        Jiaguo~Zhang \thanks{E.~Fretwurst, R.~Klanner and J.~Schwandt are with the Institute of Experimental Physics of Hamburg University, Luruper Chaussee 147, D-22761 Hamburg, Germany.}
\thanks{J.~Zhang is with DESY, Notkestrasse 85, D-22607 Hamburg, Germany.}
\thanks{I.~Kopsalis was visitor from the National Technical University, Athens, Greece, and is presently PhD student at the Institute of Experimental Physics of Hamburg University.}}
\markboth{Journal of \LaTeX\ Class Files,~Vol.~XX, No.~XX, September~2013}%
 {Shell \MakeLowercase{\textit{et al.}}: Bare Demo of IEEEtran.cls for Journals}
%



\maketitle

\begin{abstract}
The high intensity and high repetition rate of the European X-ray Free-Electron Laser, presently under construction in Hamburg, requires silicon sensors which can stand X-ray doses of up to
1~GGy for 3~years of operation at high bias voltage.
  Within the AGIPD Collaboration the X-ray-radiation damage in MOS Capacitors and Gate-Controlled Diodes fabricated by four vendors on high-ohmic n-type silicon with two crystal orientations and different technological parameters, has been studied for doses between 1~kGy and 1~GGy.
 The extracted values of oxide-charge and surface-current densities have been used in TCAD simulations, and the layout and technological parameters of the AGIPD pixel sensor optimized.
  It is found that the optimized layout for high X-ray doses is significantly different from the one for non-irradiated sensors.
 First sensors and test structures have been delivered in early 2013.
  Measurement results for X-ray doses of 0 to 10~MGy and their comparison to simulations are presented.
 They demonstrate that the optimization has been successful and that the sensors fulfill the required specifications.

\end{abstract}


%

\section{Introduction}
 \label{sect:Introduction}

The European X-ray Free-Electron Laser (XFEL) \cite{XFEL}, planned to start user operation in~2016, will provide laterally coherent X-ray pulses with a peak brilliance of 10$^{33}$ photons~/~(s~$\cdot $~mm$^2$~$ \cdot $~mrad$^2 $~$\cdot $~0.1\%~BW) and less than 100~fs duration.
  The pulses will be delivered in trains of typically 2700 pulses with more than 10$^{12}$~photons/pulse at 4.5~MHz, followed by a gap of 99.4~ms.
 The high instantaneous intensity, short pulse duration and high repetition rate pose very demanding requirements for imaging detectors \cite{Tschentscher:2011, Graafsma:2009}.

  The Adaptive Gain Integrating Pixel Detector, AGIPD \cite{AGIPD}, is currently under development to meet these challenges.
 It is a hybrid pixel-detector system with $1024 \times 1024$ $p^+$~pixels of dimensions 200~$\upmu $m $\times $~200~$\upmu $m, made of 16~$p^+ n n^+$-silicon sensors, each with 10.52~cm $\times $ 2.56~cm sensitive area and 500~$\upmu $m thickness.
  The particular detector requirements are a separation between noise and single photons down to energies of 3~keV, more than 10$^4$ photons per pixel for a pulse duration of less than 100~fs, negligible pile-up at the XFEL repetition rate of 4.5~MHz, operation for X-ray doses up to 1~GGy, good efficiency for X-rays with energies between 3 and 20~keV, and minimal inactive regions at the edges \cite{Henrich:2011}.

 In the design of the sensor the high number of photons per pixel, which cause the so-called plasma effect, and the high X-ray dose have to be taken into account.
  The plasma effect~\cite{Tove:1967} was studied in references~\cite{Becker:2010, Becker:Thesis}.
 In order to limit the charge-collection time to below 100~ns, the approximate integration time of the AGIPD readout, and to avoid too large a spread of the charges during their drift through the sensor, an operating  voltage well above 500~V is recommended.
  The sensor design presented aims at a breakdown voltage above 900~V.

  With respect to radiation damage, X-ray irradiation results in an increase of the oxide-charge density, $N_{ox}$, and the formation of traps at the Si-SiO$_2$ interface, which cause an increase of the surface-current density, $J_{surf}$.
   Results on the values of $N_{ox}$ and $J_{surf}$ as function of dose, crystal orientation, oxide thickness, and presence of an additional Si$_3$N$_4$~layer, obtained from MOS Capacitors, MOS-C, and Gate-Controlled Diodes, GCD, from four different 
vendors are presented in~\cite{Zhang:2012,Zhang:2011,Zhang1:2012,Zhang:Thesis}.

  It is found that $N_{ox}$ increases up to doses of 1 to 10~MGy and then saturates. Typical values after  annealing at 80$^\circ $C for 10~minutes are:
   $N_{ox} = 10^{12}$~cm$^{-2}$ at 10~kGy, $2\cdot 10^{12}$~cm$^{-2}$ at 1~MGy, and $3\cdot 10^{12}$~cm$^{-2}$ at 1~GGy.
   The values for the different samples differ by approximately a factor two.
  In addition, they depend on the value and the direction of a possible electric field, and annealing is observed already at room temperature.

   The values of $J_{surf}$ too, show an increase up to dose values between 1 and 10~MGy.
  For higher values the measured $J_{surf}$ values decrease, which is not yet understood.
    Typical values after annealing at 80$^\circ $C for 10 minutes are: $J_{surf} = 0.3$~$\upmu $A/cm$^2$ at 1~kGy, and 2 to 6~$\upmu $A/cm$^2$ for the maximal values.
  Again, the values differ significantly for the different technologies. They also depend on a possible electric field, and annealing is observed.

 For $p^+ n n^+$ sensors the main impact of $N_{ox}$ are high electrical fields at the edges of the depletion boundaries of the $p^+$ implants close to the Si-SiO$_2$ interface, which reduce the breakdown voltage, $V_{bd}$.
  In addition, an electron-accumulation layer forms at the Si-SiO$_2$ interface, which changes the electric field and the charge-collection efficiency in this region~\cite{Poehlsen:2012}.

 The main challenge for the sensor design is achieving a high breakdown voltage, $V_{bd}$, for a highly non-uniform dose distribution in the range between a few kGy and 1~GGy.
  Extensive simulation with SYNOPSYS TCAD \cite{Synopsis} using the experimentally determined values of $N_{ox}$ and $J_{surf}$ have been used for the  optimization of the pixel and guard-ring layout.
 Preliminary results, which show that the specifications for the AGIPD sensor can be achieved, can be found in \cite{Schwandt:2012,Schwandt:2013}.

  This paper presents detailed information on these simulations and the expected sensor performance parameters.
 Based on these simulations, sensors and test structures had been ordered in autumn 2012 at SINTEF~\cite{Sintef}.
  The first batch has been delivered in early February 2013.
 Detailed measurements on both sensors and test structures before and after X-ray irradiation up to dose values of 10~MGy have been performed since.
  The results and comparisons to the simulations are presented in this paper.
 They show that the optimization has been successful:
  The breakdown voltage, $V_{bd}$, exceeds 900~V for the entire dose range, and parameters like dark currents and inter-pixel capacitances are well within specifications.

 \section{Sensor Optimization}
  \label{sect:SensorOpt}
We first present the specification for the AGIPD sensors, and then give details on the simulations, in particular how the X-ray-radiation damage has been implemented.
  Finally the optimization for both guard rings and pixels will be discussed. \\

   \subsection{Sensor Specifications}
  Table~\ref{tab:Specs} presents the main specifications for the AGIPD sensors.
 In addition to the geometry specifications, separate electrical specifications before irradiation and for the entire X-ray-dose range from 0 to 1~GGy are given.
  The specifications for the non-irradiated sensor are used to assure that basically the produced sensors are functioning properly.
 The specifications as function of X-ray dose will only be verified for a sample of test sensors  with $7 \times 7$ pixels from a selected set of production batches.
  In addition, test structures like, pad diodes, MOS capacitors, gate-controlled diodes, and strip sensors are available to verify technological parameters, and determine their dependence on X-ray dose.
 \begin {table}[htb]
  \centering
   \caption{Specifications for the AGIPD sensor.}
   \begin{tabular}{@{}ll@{}}
   \toprule
    \textbf{Parameter} & \textbf{Value} \\
   \midrule
    \multicolumn{2}{c}{\textsl{Geometry parameters}} \\ \midrule
    Sensitive area & $10.52$~cm~$\times $~2.56~cm \\
    Thickness & $500 \pm 20 $~$\upmu$m \\
    Pixel dimensions & $200 $~$\upmu$m $\times $~$200$~$\upmu$m \\
    Deviation from flatness & $< 20 $~$\upmu$m \\
    Distance pixel to cut edge & $1200 \pm 5 $~$\upmu$m \\ \addlinespace
  \multicolumn{2}{c}{\textsl{Electrical parameters before irradiation}} \\ \midrule
   $n$ doping of Si crystal& $3-8$~k$\Upomega\cdot $cm \\
   Doping non-uniformity & $ < 10 \% $ \\
   Breakdown voltage & $ > 900$~V \\
   Dark current of all pixels@500V & $ < 200$~nA \\
   Dark current of single pixel@500V & $ < 20$~nA  \\
   Dark current of CCR@500V & $ < 200$~nA  \\ \addlinespace
 \multicolumn{2}{c}{\textsl{Electrical parameters for 0 Gy to 1 GGy }} \\ \midrule
   Breakdown voltage & $ > 900$~V  \\
   Dark current of all pixels@500V & $ < 50 $~$\upmu$A  \\
   Dark current of single pixel@500V & $ < 50$~nA  \\
   Dark current of CCR@500V & $ 20 $~$\upmu$A  \\
   Inter-pixel capacitance & 500~fF  \\
  \bottomrule
  \end{tabular}
  \label{tab:Specs}
 \end{table}

\subsection{TCAD Simulations}
 For the optimization of the sensor, process and device simulations have been performed.
  The doping profiles of the $p^+$~implants were obtained from 2D~process simulations of the boron implantation of approximately $10^{15}$~cm$^{-2}$ and the implant annealing step for the process of one vendor~\cite{Schwandt:2012}.
 To verify the correctness of the simulation, the simulated boron depth profile was compared to results of SIMS (Secondary Ion Mass Spectroscopy) measurements for the same process.
  For a phosphorus bulk doping of $10^{12}\,$cm$^{-3}$ and $<$111$>$ crystal orientation, the resulting junction depth is $1.2\,\upmu$m with a lateral extension of $1.0\,\upmu$m, measured from the edge of the implant window.
 In addition, simulations of different processes resulting in deeper implants have been performed.
  Finally, for the optimized design a doping  process  which results in a junction depth of $2.4\,\upmu$m and a lateral extension of $1.95\,\upmu$m has been used.

 The electrical device simulations were performed for a temperature of 293~K using the drift-diffusion model.
  The carrier lifetime in the bulk was assumed to be 1~ms, the mobility was modeled doping dependent with a degradation at the interface, and for the avalanche process the van Overstraeten -- de Man model with the default parameters~\cite{Synopsis} was used.

 The X-ray-radiation damage was implemented in the following way: The oxide charge density, $N_{ox}$, was simulated as a charge-sheet located at the Si-SiO$_2$ interface with a uniform distribution along the interface.
  For the simulation of the surface-current density, $J_{surf}$, it was assumed, that the surface current is generated via a trap at mid-gap with the same surface recombination velocity, $s_r$, for electrons and holes.
 In table~\ref{tab:Radpar} the values, which were used in the simulations, for $N_{ox}$, $J_{surf}$ and $s_r$ are given.
  The corresponding dose values were estimated from measurements on X-ray-irradiated MOS capacitors and gate-controlled diodes after annealing at 80$^\circ $C for 10~minutes~\cite{Zhang1:2012}.
 As it has been found, that $N_{ox}$ and $J_{surf}$ saturate or even decrease for dose values above approximately 10~MGy, the maximum values given in table~\ref{tab:Radpar} are considered valid for the entire dose range between 10~MGy and 1~GGy, the highest values expected at the European XFEL.
\begin{table}[htb]
  \centering
  \caption{X-ray radiation damage parameters used in the simulations, and the corresponding dose values.}
  \begin{tabular}{@{}cccc@{}}
  \toprule
  Dose &  $N_{ox}$ & $J_{surf}$ & $s_{r}$ \\ 
  $\lbrack$MGy$\rbrack$ &  $\lbrack$cm$^{-2}\rbrack$  & $\lbrack\upmu$A/cm$^2\rbrack$ & $\lbrack$cm/s$\rbrack$ \\
 \cmidrule(r){1-1} \cmidrule(l){2-4}
  0 &  $1\cdot 10^{11}$ & 0.005 & 8 \\
  0.01 &  $1\cdot 10^{12}$ & 2.3 & 3580 \\
  1 &  $2\cdot 10^{12}$ & 5 & 7500 \\
  10 -- 1000&  $3\cdot 10^{12}$ & 8 & 12040 \\
  \bottomrule
  \end{tabular}
  \label{tab:Radpar}
\end{table}

 On top of the SiO$_{2}$ Neumann boundary conditions (zero normal component of the electric field) were used.
 The results from references~\cite{Poehlsen:2012, Poehlsen:2013, Poehlsen:Thesis} imply, that the boundary conditions on top of the oxide separating $p^+$ implants at the same potential change with time from Neumann to Dirichlet (constant potential). The corresponding time constants vary between minutes and days, depending on environmental conditions like relative humidity.
  We have decided to use Neumann boundary conditions, as simulations have shown, that they result in a lower breakdown voltage after X-ray irradiation.
 They are also considered to be valid for the sensor operation in vacuum.
  No additional passivation layer on top of the SiO$_{2}$ has been simulated.

 For the simulations we typically used 16 threads of Intel Xeon E5520 2.5~GHz PCs.
  For the 2-D pixel simulations the minimum mesh size  close to the Si-SiO$_2$ interface and at the edge of the $p^+$ implantation required to obtain reliable results, was 3~nm transverse and 6~nm perpendicular to the interface.
 The total number of vertices was 135~000, and the number of elements 255~000.
  An $I-V$ scan from 0 to 1000~V in 10~V steps took 30~minutes for an oxide-charge density of $10^{11}$~cm$^{-2}$ and 70~minutes for $3 \cdot 10^{12}$~cm$^{-2}$.
 For the 3-D simulation of a quarter of a pixel, the minimal mesh size used was  $30 \times 90 \times 90 $~nm$^3$, resulting in 900~000 vertices and 5~000~000 elements.
  A voltage scan took between 30 and 200~hours, and the iterative solver required approximately 80~GB of main memory.

\begin{figure}[htb]
  \centering
  \includegraphics[width=1\linewidth]{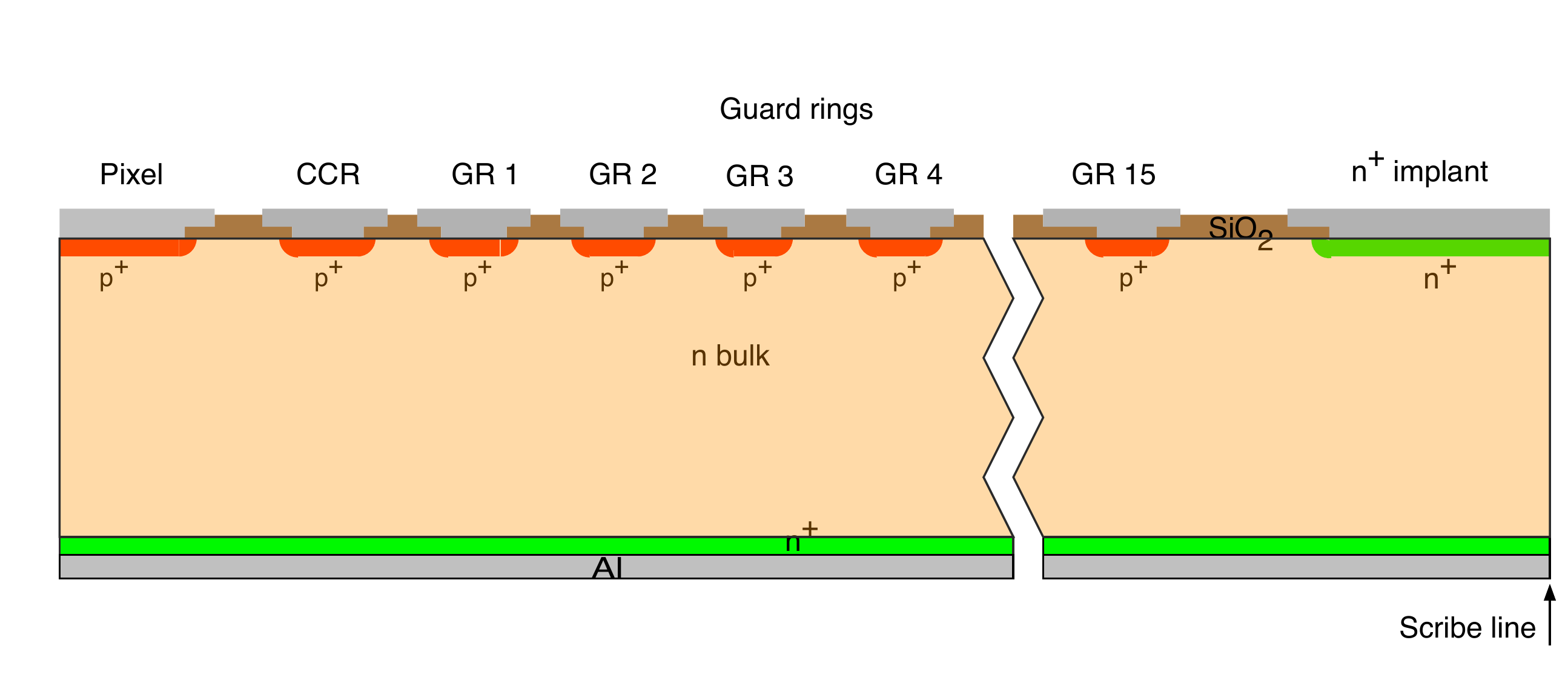}
  \caption{Layout of the sensor region simulated for the guard ring optimization.}
  \label{fig:guardsketch}
\end{figure}

\subsection{Simulation of the Guard Rings}

 In order to optimize the design of the pixel and the guard-ring structure for high breakdown voltage, $V_{bd}$, several parameters have been varied in the simulations.
  As criterium for $V_{bd}$ we used the quantity $K = (dI/dV)/(I/V)$ \cite{Bacchetta:2001} with $K_{bd} = 10$.
 The value $K= 1$ corresponds to an ohmic resistor, and $K\gg 1$ to avalanche breakdown.
  The results obtained are insensitive to the exact choice of the value for $K_{bd}$.

 The basic layout simulated is shown in figure~\ref{fig:guardsketch}: A strip with the width of half a pixel, $n_{guard}$~guard rings, a current collection ring, $CCR$, and an $n^+$ scribe-line implant.

 As  first step, the optimal junction depth, $d_{jun}$, and oxide thickness, $t_{ox}$, were determined for a structure with $n_{guard} = 0$ (0~GR).
  The breakdown voltage, $V_{bd}$, has been determined as function of $t_{ox}$ for different values of the bulk doping, of $N_{ox}$, $d_{jun}$ and the Al overhang of the CCR towards the $n^+$ scribe-line implant.
 For the Al overhang towards the pixels a value of $5\;\upmu$m has been chosen.

\begin{figure}[htb]
  \centering
  \includegraphics{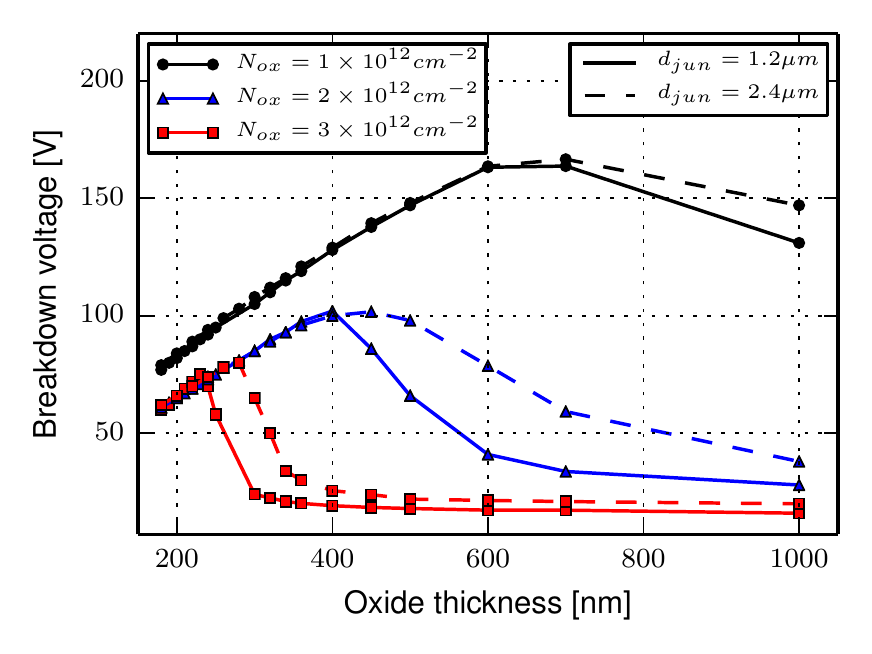}
  \caption{Breakdown voltage for 0 GR (CCR only) as function of the oxide thickness for different values of $N_{ox}$, a Si-bulk resistivity of 5~k$\Omega\cdot$cm, an Al overhang of 5~$\upmu$m on both sides, and junction depths $d_{jun}$ of 1.2~$\upmu$m (solid) and 2.4~$\upmu$m (dashed). The breakdown voltages for $N_{ox}$ values below $10^{12}$~cm$^{-2}$ are above 180~V (not shown).}
  \label{fig:Vbd_vs_tox}
\end{figure}

 The results for a bulk doping of $8.8\cdot 10^{11}\;cm^{-3}$ (resistivity of 5~k$\Omega\cdot$cm) and an Al overhang of 5~$\upmu$m for the CCR, are shown in figure~\ref{fig:Vbd_vs_tox}.
  For a given $N_{ox}$, $V_{bd}$ as function of $t_{ox}$ has a maximum, which we call optimum $t_{ox}$.
 The optimum $t_{ox}$ decreases with increasing $N_{ox}$.
  For $N_{ox} = 1\cdot 10^{12}$~cm$^{-2}$ the optimum $t_{ox}$ is between 600 and 700~nm, depending on $d_{jun}$, which corresponds to the optimization for little or no X-ray radiation damage.
 For the assumed saturation value of $N_{ox} = 3\cdot 10^{12}$~cm$^{-2}$, the optimum values are 230~nm and 270~nm for $d_{jun} = 1.2\;\upmu$m and  $d_{jun} = 2.4\;\upmu$m, respectively, with values for $V_{bd}$ between 70 and 80~V.
  After reaching its maximum, $V_{bd}$ decreases rapidly with increasing $t_{ox}$ to values below 25~V, which corresponds to the breakdown voltage of a structure without Al overhang of the CCR.
 This sudden decrease of $V_{bd}$ is related to the electron-accumulation layer below the Al overhang~\cite{Schwandt1:2012}:
  If an accumulation layer is present, there is a single high-field region at the corner of the $p^+n$~junction.
 Without accumulation layer the voltage drop occurs over the entire region below the Al layer, with one high field region at the edge of the junction and a second one below the end of the Al overhang.
  In addition, a deeper $p^+n$~junction results in a lower field at the edge of the junction, thus further increasing the breakdown voltage.

 Based on these results and on 2D simulations for the pixel, an oxide thickness of 250~nm and a 2.4~$\upmu$m deep $p^+$~implant have been chosen for the further optimization as compromise between technological feasibility and high $V_{bd}$.
  From the value of $V_{bd} \approx $~70~V  with zero guard rings we estimated that 15~floating guard rings, with outward Al overhangs of 5~$\upmu$m, would be required to reach $V_{bd}$ values close to 1000~V.

 As next step, 2D simulations in Cartesian coordinates for the optimization of the straight sections of the guard rings have been performed: The spacing between the guard rings, their implant widths and inward (towards the pixels) Al overhangs were varied until the voltage differences between adjacent guard rings were approximately the same for the saturation value of the oxide-charge density of $3\cdot 10^{12}$~cm$^{-2}$.
  Finally, simulations for the corners of the guard-ring structure were performed in cylindrical coordinates, in order to check that  $V_{bd}$ does not decrease significantly at the corners.
 Cylindrical symmetry around the center of the circle defining the outer pixel corner has been assumed.

 The optimized design parameters for the guard-ring structure are given in table~\ref{tab:Designpar}.
  In table~\ref{tab:breakdownvsdose} the results of the 2D simulations in Cartesian (2D~(x,~y)) and cylindrical (2D~(r,~z)) coordinates of the breakdown voltage as function of oxide-charge density and bulk resistivity for the optimized guard-ring design are given.
 They show that the AGIPD specifications for the breakdown voltage can be met over the entire dose range.

\begin{table}[htb]
  \centering
  \caption{Optimized design parameters for the AGIPD sensor.}
   \begin{tabular}{@{}ll@{}}
   \toprule
   \textbf{Parameter} & \textbf{Value} \\
   \midrule
   \multicolumn{2}{c}{\textsl{Technological parameters}} \\ \midrule
    Oxide thickness & 250~nm \\
    Junction depth & 2.4 $\upmu$m \\
  \multicolumn{2}{c}{\textsl{Guard rings}} \\ \midrule
  Gap between pixel- and CCR $p^+$ implants& 20 $\upmu$m \\
  Width of the $p^+n$~junction of the CCR & 90 $\upmu$m \\
  Al overhang of the CCR & 5 $\upmu$m \\
  Gap between CCR and GR 1 & 12~$\upmu$m \\
  Widths of the $p^+n$ junctions of the GRs & 25~$\upmu$m \\
  Al overhangs towards pixels for GRs 1-15 & 2, 3, 4, ..., 16 $\upmu$m \\
  Al overhangs towards $n^+$ implant for GRs 1-15 & 5 $\upmu$m \\
  Gaps between GRs 1-2, 2-3,..., 14-15 & 13.5, 15, 16.5, ..., 33 $\upmu$m \\
  Distance GR 15 to $n^+$ scribe-line implant & 50 $\upmu$m \\
  Width of $n^+$ scribe-line implant & 340 $\upmu$m \\
  \multicolumn{2}{c}{\textsl{Pixel}} \\ \midrule
  Gap between pixel $p^+$ implants& 20 $\upmu$m \\
  Al overhang & 5 $\upmu$m \\
  Radius of implant at the corners  & 10 $\upmu$m \\
  Radius of Al at the corners  & 12 $\upmu$m \\
  \bottomrule
  \end{tabular}
  \label{tab:Designpar}
\end{table}
\begin{table}[htb]
	\centering
     \caption{Breakdown voltages obtained from the 2D simulations in Cartesian and cylindrical coordinates as function of oxide-charge density and bulk resistivity for the optimized guard-ring structure.}
		\begin{tabular}{@{}lrrrr@{}}
		\toprule
		     & \multicolumn{2}{c}{5 k$\Omega\cdot$cm}
		     & \multicolumn{2}{c}{8 k$\Omega\cdot$cm}  \\
		      \cmidrule(lr){2-3} \cmidrule(l){4-5}
		    $N_{ox}$ [cm$^{-2}$] & 2D (x, y) & 2D (r, z) &
		                         2D (x, y) & 2D (r, z) \\
		     \cmidrule(r){1-1} \cmidrule(l){2-5}
		    $1\cdot10^{12}$ & $>1100$ V & $>1100$ V &
		                        $>1100$ V & $>1100$ V \\		    $2\cdot10^{12}$ & 1080 V &   910 V &
		                          950 V &   950 V \\
		    $3\cdot10^{12}$ &  $>1100$ V & 910 V &
		                         1000 V & 960 V \\
		\bottomrule
		\end{tabular}
		\label{tab:breakdownvsdose}
\end{table}

\begin{figure}[htb]
  \centering
  \includegraphics[width=0.45\linewidth]{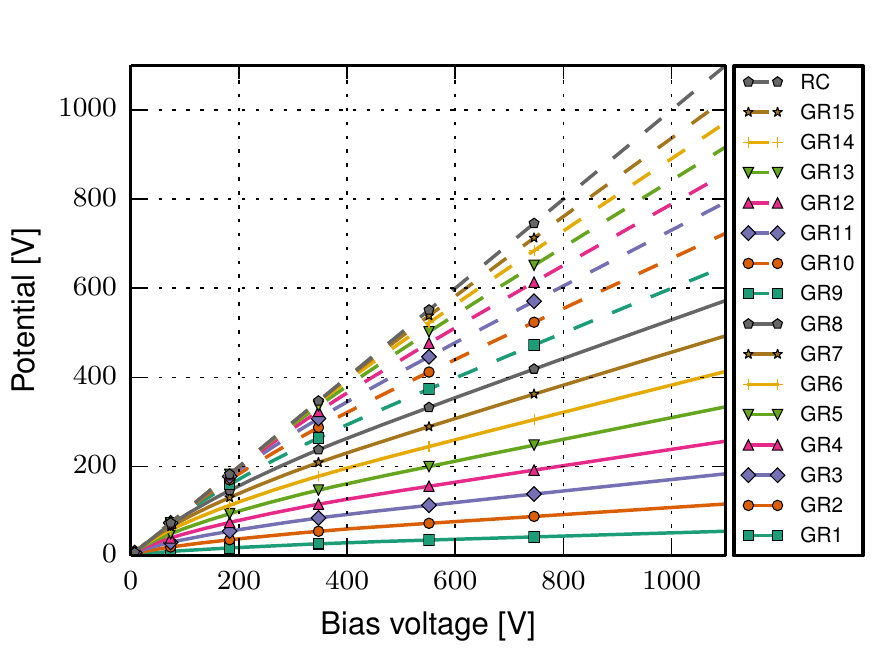}
    \includegraphics[width=0.45\linewidth]{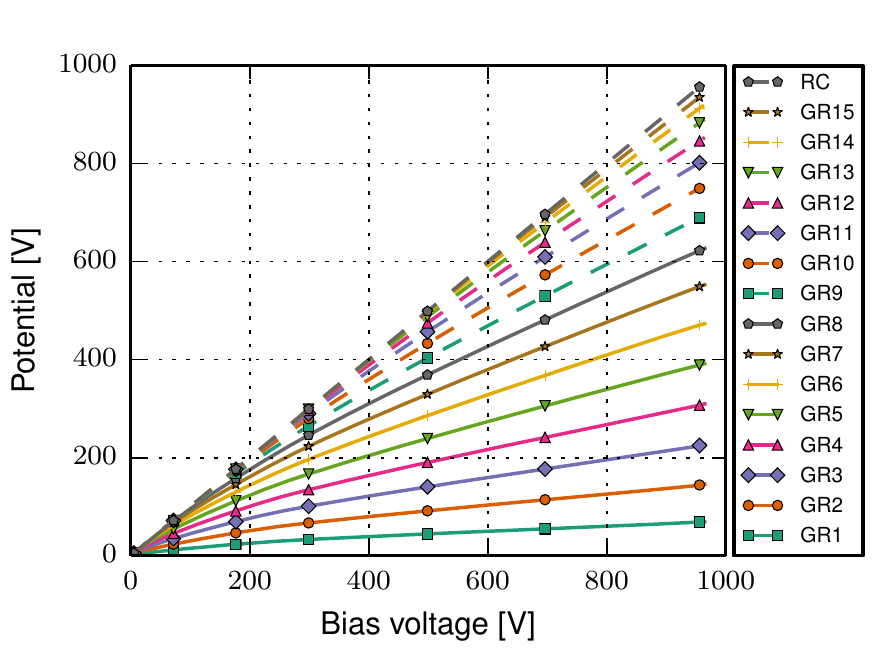}
  \caption{Simulated voltages of the 15 guard rings (GR) vs. bias voltage for $N_{ox} = 3\cdot 10^{12}$~cm$^{-2}$ and a resistivity of 5~k$\Omega\cdot$cm. The voltage is applied at the rear contact (RC). Top: Simulation of the straight section of the guard rings in Cartesian coordinates. Bottom: Simulation of the corners in cylindrical coordinates.}
  \label{fig:pot_gr15}
\end{figure}

 The difficulties in determining the optimal spacing between the implants of the guard rings for high breakdown voltage at the straight sections as well as  at the corners of the guard-rings, can be seen in figure~\ref{fig:pot_gr15}, which shows the simulated guard-ring voltages vs.\ bias voltage in Cartesian (top) and cylindrical coordinates (bottom)  for $N_{ox} = 3\cdot 10^{12}$~cm$^{-2}$ and a bulk resistivity of 5~k$\Omega\cdot$cm.
  For the Cartesian case the voltage differences between adjacent guard rings are similar at high voltages, the contributions of the individual guard rings to the overall voltage drop are similar and a high breakdown voltage with values above 1100~V is achieved.
   In cylindrical coordinates the individual guard rings are on a higher potential compared to the Cartesian case, and the voltage drop between the inner guard rings is higher than between the outer ones.
 As a result, the breakdown voltage is reduced to 910~V.
  The breakdown voltage at the corners could be increased by decreasing the spacing between the inner guard rings.
 This however, would violate the design rules for the first ring.
  A reduction of the implant width of the guard rings, e.\ g.\ to 15~$\upmu$m, while keeping the other parameters the same, would result in a further decrease of the breakdown voltage to 765~V.

  Next we discuss the breakdown which occurs, when the depletion zone in the bulk reaches the cut edge.
 The 340~$\upmu$m wide $n^+$ scribe-line implant is introduced to prevent that this happens at too low voltages.

  Figure~\ref{fig:pot_gr15} shows, that the last guard ring, Gr15, is not at the same potential as the $n^+$ scribe-line implant which, via the cut edge and the non-depleted bulk, is at the rear-contact potential.
   For high oxide-charge densities there is an accumulation layer at the Si-SiO$_2$ interface, and the rear-contact potential nearly reaches the last guard ring.
 There, in the silicon bulk, the depletion zone ends and the non-depleted region starts.
  If the oxide-charge density is too low to cause an accumulation layer, the edge of the depletion zone will move to the edge of the $n^+$ implant.
 In this case, the depleted region will reach the cut edge at a lower voltage, than for the case of high oxide-charge densities.
  The simulations in \cite{Schwandt:2013} show, that at 880~V for $N_{ox} = 5 \cdot 10^{10}$~cm$^{-2}$ and a bulk resistivity of 8~k$\Omega\cdot$cm, the minimal distance between the edge of the depletion zone and the cut edge is 50~$\upmu$m, whereas it is  25~$\upmu$m for $N_{ox} = 1 \cdot 10^{10}$~cm$^{-2}$.
 This distance also increases if the resistivity of the silicon crystal is reduced: For an oxide-charge density $N_{ox} = 5 \cdot 10^{10}$~cm$^{-2}$ and a resistivity of 5~k$\Omega\cdot$cm it is 65~$\upmu$m.
  We consider a distance of 25~$\upmu$m reasonably save with respect to damages close to the cut edge.

\subsection{Simulation of the Pixels}

\begin{figure}[htb]
  \centering
  \includegraphics[width=0.75\linewidth]{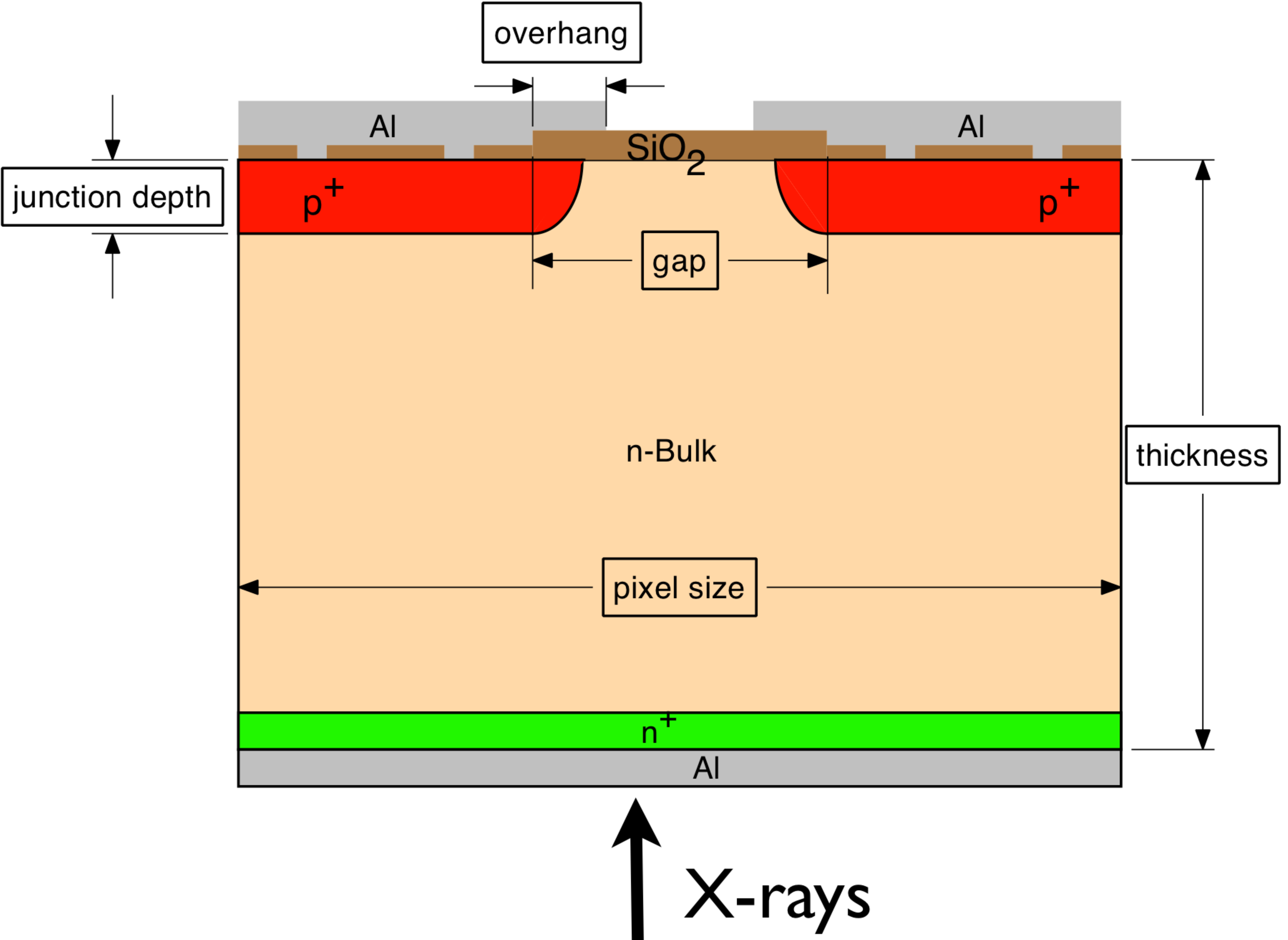}
  \caption{Layout of the sensor region simulated for the pixel optimization.}
  \label{fig:pixelsketch}
\end{figure}

 For the design of the pixels the following parameters, most of them shown in~figure~\ref{fig:pixelsketch}, have been optimized: Junction depth, oxide thickness, gap between the $p^+$~implants, Al overhang and radii of the implant window, and Al layer at the corners.
  These parameter have been optimized with respect to  breakdown voltage, dark current and inter-pixel capacitance.
 Although a square pixel is inherently a 3D problem, 2D strip simulations were used for the optimization.
  The reason is that due to the large pixel size and the fine mesh at the interface, which is required to describe accurately the surface damage, the computer resources needed for the optimization were too large.
 Instead, the 2D results for the dark current and inter-pixel capacitance were extrapolated to 3D values, and finally the breakdown voltage and the dark current were checked with a 3D simulation of a quarter of a pixel.

 In \cite{Schwandt:2012} it was shown that a junction depth of 1.2~$\upmu$m is not sufficient to reach a breakdown voltage above 900~V for the pixels.
  However, it could be shown, that for a junction depth of 2.4~$\upmu$m and an oxide thickness of 300~nm the breakdown voltage exceeds 1000~V for 20--40~$\upmu$m wide gaps and a metal overhang of 5~$\upmu$m.
 We finally choose a gap of 20~$\upmu$m to minimize possible effects from time-dependent boundary conditions on the SiO$_2$ surface and also to minimize the area where surface currents can be generated.
  Finally, the 3D simulation of a quarter of a pixel for the final design with a 250~nm thick oxide predicts, that the optimized design should meet all AGIPD specifications for the required dose range from 0 to 1~GGy~\cite{Schwandt1:2012}.

\section{Measurement Results and Comparison to Simulations}

The AGIPD sensors were fabricated by SINTEF on 6~inch wafers with $<$100$>$ crystal orientation and resistivities between 7.4 and 8.6~k$\Omega\cdot$cm.
  To achieve the desired junction depth and oxide thickness, the standard process was modified.
 For the  $p^+$-implants the same technological parameters as for the process simulation have been used. The junction depth was checked by spreading-resistance measurements.

 For one wafer of the first delivery,  sensor-acceptance and radiation-hardness measurements have been made to verify the design.
  The X-ray irradiations were done at the beam line P11 of PETRA III with monochromatic 12~keV X-rays for doses up to 1~MGy, and with 8~keV X-rays up to 10~MGy.
 The dose rate was between 0.42 and 7.1~kGy/s.
  During the irradiations the contacts of the test structures and sensors were kept floating.
 The measurements were performed within one hour after the irradiation, and after annealing for 10~minutes at 80$^\circ$C. \\

\subsection{Results from the Test Structures}

 The test structures investigated were a circular MOS-capacitor (MOS-C) with a diameter of 1.5~mm,  a Gate-Controlled Diode (GCD) with a finger structure with 100~$\upmu$m wide gates and 100~$\upmu$m wide diodes, and a square $p^+n$~diode of 5.05~mm~$ \times $~5.05~mm.
 From the $C-V$ measurements of the $p^+n$~diode a bulk resistivity of 7.8~k$\Omega\cdot$cm, and from the MOS-C capacitance in accumulation an oxide thickness of 245~nm were determined.

\begin{figure}[htb]
  \centering
  \includegraphics{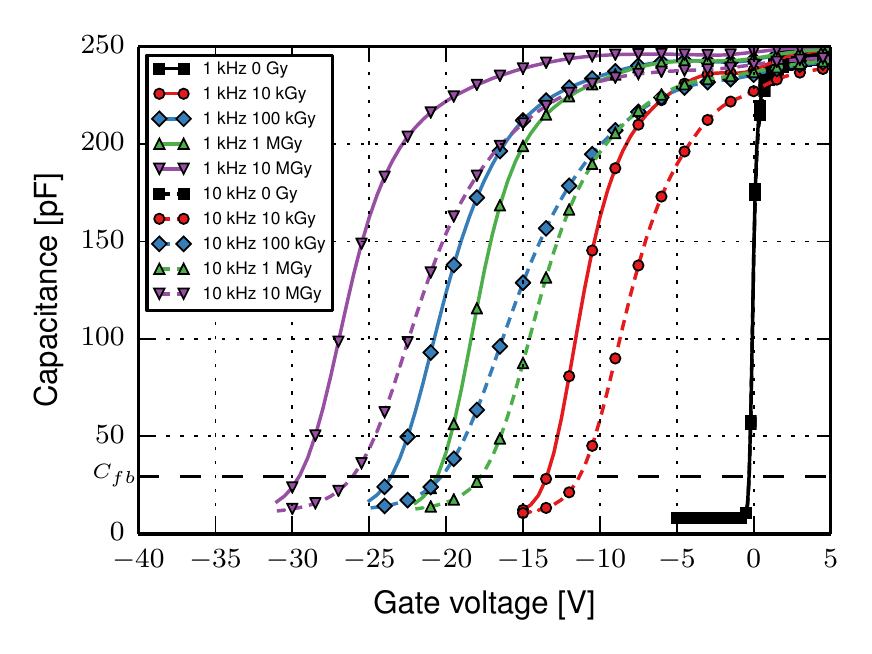}
  \caption{$C-V$ curves for a non-irradiated and an irradiated MOS-C produced together with the AGIPD sensors after annealing for 10 minutes at 80$^\circ$C. The value of the flatband capacitance, $C_{fb}$, is indicated by a horizontal line.
   Due to problems during the irradiation, the 10~MGy value may be somewhat lower.}
  \label{fig:MOS_CV}
\end{figure}

 Figure~\ref{fig:MOS_CV} shows the results of the $C-V$ measurements at 1~and 10~kHz for the MOS-C before irradiation, and after irradiation to 0.01, 0.1, 1 and 10~MGy after annealing for 10~minutes at 80$^\circ$C.
  We only present the data for which the voltage was ramped from positive values (accumulation) to negative values (inversion).
 As already observed in~\cite{Zhang:2012}, the irradiated MOS-Cs show  hysteresis effects, which however, we do not discuss further here.

  Before irradiation the flatband voltage is $V_{fb} = -0.39$~V (flatband capacitance $C_{fb} = 29.3$~pF).
 Considering that the metal-semiconductor work-function difference $\Phi_{MS} = -0.42$~V, we conclude that the effective oxide-charge density is close to zero, probably even negative.
  In previously investigated samples from SINTEF with an oxide thickness of 750~nm, only positive oxide-charge densities had been observed~\cite{Zhang:Thesis}.
 The difference is probably related to changes in the production process which attempted to minimize the oxide-charge density before X-ray irradiation.

 The $C-V$~curves of the non-irradiated MOS-C are similar to the one of an ideal MOS-C~\cite{Grove:1967}, and show no frequency dependence.
  The $C-V$~curves for the irradiated MOS-C show large shifts to negative voltages, a stretching of the transition from accumulation to inversion, and a strong dependence on frequency, as expected for radiation-induced positive oxide charges and interface traps.
 To extract the oxide-charge density, $N_{ox}$, as function of dose, and to compare the results to previous measurements \cite{Zhang:Thesis}, the flatband-voltage shift of the 1~kHz curves was used.
  In table~\ref{tab:nox_jsurf} the values of $N_{ox}$ for the different dose values, directly after irradiation and after annealing for 10 minutes at 80$^\circ$C, are given.
 The values agree with the previous measurements on SINTEF samples with 750~nm oxide, which showed that a saturation value of $2-4 \cdot 10^{12}$~cm$^{-2}$ is reached for a dose of about 100~kGy.

  From the $I-V$~measurement on a GCD, a surface-current density of $J_{surf} = 1.9$~nA/cm$^2$ at $T=20^\circ$C before irradiation is extracted.
 This is four times lower than the value found previously for test structures from SINTEF~\cite{Zhang:Thesis}.
  The values for $J_{surf}$ after irradiation and  annealing for 10 minutes at 80$^\circ$C are given in table~\ref{tab:nox_jsurf}.
  A saturation value of about 1.5~$\upmu $A/cm$^2$ at a temperature of $20^\circ $C is reached for a dose of about 1~MGy.

 To summarize: Apart from a small negative oxide-charge density before irradiation, the values of the oxide-charge and surface-current densities before and after X-ray irradiation determined from the test structures produced by SINTEF on the same wafers as the AGIPD sensors are similar to the ones from the test samples from the standard SINTEF process, which we obtained in autumn 2012.
  The values are somewhat different from the ones shown in table~\ref{tab:Radpar}, which were used for the sensor optimization.
 The latter have been determined until spring 2012 from test structures from three other vendors~\cite{Zhang:Thesis}.
  The observed difference do not invalidate the optimization of the AGIPD-sensor design.
   However, they result in some differences between the results of the simulations and the measurements.

\begin{table}[htb]
	\centering
     \caption{Dependence on X-ray dose of the oxide-charge density $N_{ox}$ and the surface-current density $J_{surf}$ at $20^\circ$C from test structures produced together with the AGIPD sensors.}
		\begin{tabular}{@{}lcccc@{}}
		\toprule
		     & \multicolumn{2}{c}{As irradiated}
		     & \multicolumn{2}{c}{After 10~min.@80$^\circ$C}  \\
		      \cmidrule(lr){2-3} \cmidrule(l){4-5}
		 Dose & $N_{ox}$  & $J_{surf}$ & $ N_{ox}$ & $J_{surf}$ \\
$\lbrack$MGy$\rbrack$ & $\lbrack$cm$^{-2}\rbrack$ & $\lbrack\upmu$A/cm$^2\rbrack$ & $\lbrack$cm$^{-2}\rbrack$ & $\lbrack\upmu$A/cm$^2\rbrack$ \\
		     \cmidrule(r){1-1} \cmidrule(l){2-5}
          0.01 &  $1.8\cdot10^{12}$ & 0.3  & $1.2\cdot 10^{12}$ & 0.3\\ 		 0.1  & $3.6\cdot10^{12}$  & 1.5 & $2.1\cdot10^{12}$ &  1.2\\ 		   1   &  $3.0\cdot10^{12}$ & 2.7 & $1.8\cdot10^{12}$ & 1.6 \\
		  10   &  $3.4\cdot10^{12}$ & 1.9 & $2.6\cdot10^{12}$ & 1.5 \\
		\bottomrule
		\end{tabular}
		\label{tab:nox_jsurf}
\end{table}

\subsection{Results for the Sensors}

\begin{figure}[htb]
  \centering
  \includegraphics[width=0.5\linewidth]{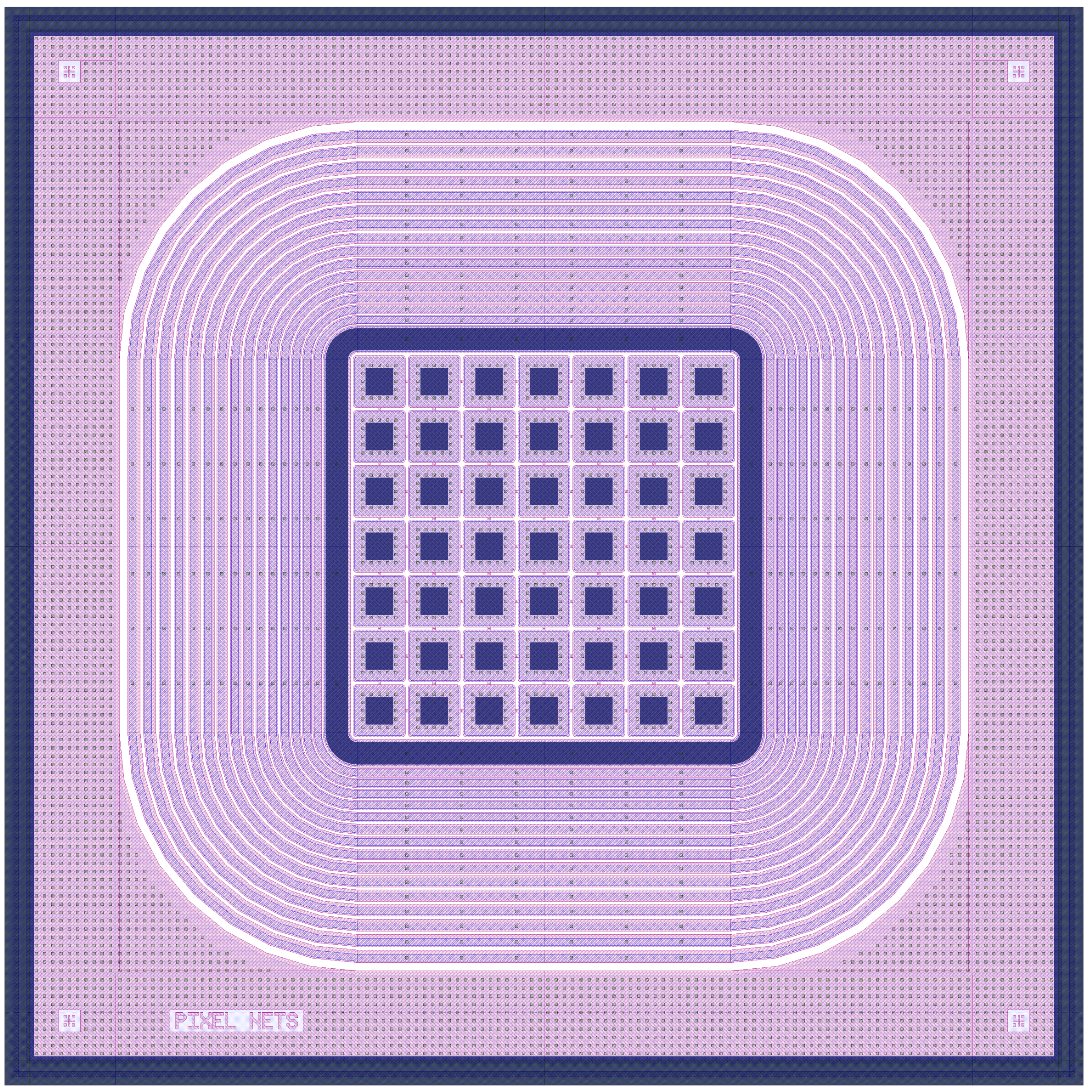}
  \caption{Layout of the 7$\times$7 test sensor with all pixels interconnected.}
  \label{fig:7x7_gds}
\end{figure}

 For the evaluation of the performance of the pixels and of the guard-ring structure two types of test sensors with 7$\times$7 pixels surrounded by the optimized guard rings were fabricated (see figure~\ref{fig:7x7_gds}).
  In the first type, which was used for the $I-V$ measurements, all pixels are connected via narrow Al bridges to their direct neighbors.
 In the second type, which was used for the inter-pixel capacitance measurements, the center pixel is surrounded by 3 rings of connected pixels.
  The irradiations were carried out in two ways: In the first the sensor was irradiated uniformly; in the second, using a 1~mm thick Ta absorber, only half of the sensor was irradiated, in order to produce a highly non-uniform irradiation.

\begin{figure}[htb]
  \centering
  \includegraphics[width=0.49\linewidth]{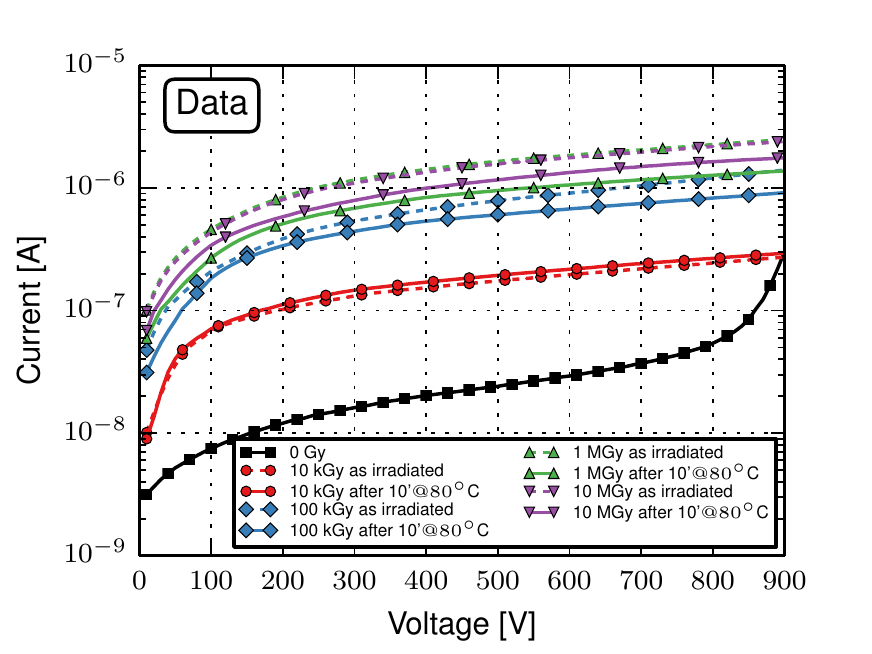}
    \includegraphics[width=0.49\linewidth]{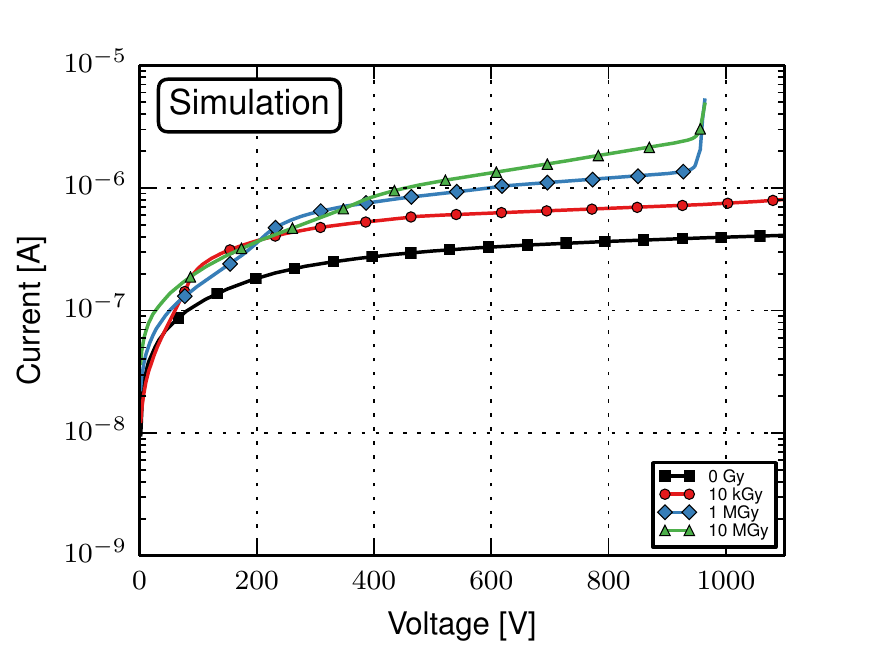}
  \caption{Top: CCR current of the 7$\times$7 test sensor measured at 20$^\circ$C and scaled to the full AGIPD sensor for  different dose values before and after annealing.
  Bottom: Simulations of the CCR current  scaled to the full AGIPD sensor for the different dose values, with the values of $N_{ox}$ and $J_{surf}$ used for the optimization and given in table~\ref{tab:Radpar}.}
  \label{fig:CCR_IV}
\end{figure}

 The top figure~\ref{fig:CCR_IV} shows as function of voltage the measured CCR currents, scaled to the full size of the AGIPD sensor, for the non-irradiated and the uniformly irradiated sensor at 20$^\circ$C in a dry (relative humidity below 5~\%) atmosphere.
  The measurements have been done within one hour after the irradiation, and in addition, after annealing for 10~minutes at 80$^\circ$C.
 The maximum voltage of 900~V is due to limitations of the cold chuck of the probe station.

 We note, that the non-irradiated sensor shows a soft breakdown at approximately 800~V.
  The reason was already explained in section~\ref{sect:SensorOpt}: The low oxide-charge density at the Si-SiO$_2$ interface and the high bulk resistivity causes the depletion region to touch the cut edge.
 This explanation is confirmed by the observations, that the soft breakdown is absent before cutting the wafer and that it disappears after a low dose irradiation of 200~Gy.
  This soft breakdown does not cause a problem for the actual use of the AGIPD sensor, as already during the calibration they will be exposed to an X-ray dose in excess of 200~Gy.

 For the irradiated sensors no breakdown up to 900~V is observed.
  The current increases with dose and saturates around 1~MGy, compatible with the surface-current densities shown in table~\ref{tab:nox_jsurf}.
 The maximal current measured at 20$^\circ$C and 900~V after annealing for 10~minutes at 80$^\circ$C is about 2~$\upmu$A.
  The value at --20$^\circ$C is  40~nA.
 This annealing reduces the currents by about a factor 2.
  All values are well within the specifications given in table~\ref{tab:Specs}.
 We note here, that a sensor not optimized for X-ray radiation hardness with 12 guard rings produced by SINTEF after irradiation to 100~kGy has a breakdown voltage of about 300~V~\cite{Schwandt:2013}, which demonstrates that the optimization has been necessary and successful.

 The bottom figure~\ref{fig:CCR_IV} shows for comparison the predictions from the simulations for the CCR currents with the parameters given in section~\ref{sect:SensorOpt}.
  For the predictions of the full-size sensor, the CCR currents from the 2D simulation in Cartesian coordinates were scaled to the length of the straight sections of the AGIPD guard ring and the currents from the 2D simulation in cylindrical coordinates added.
 The breakdown at 910~V for oxide-charge densities above $10^{12}$~cm$^{-2}$ is due to the breakdown at the corners of the CCR.
  No such breakdown is observed in the measurements, showing that the assumptions made in the simulations have been conservative.
 We note, that for the irradiated sensor the predicted and measured currents agree.
  For the non-irradiated sensor, the apparent disagreement is due to the assumed generation lifetime of 1~ms.

\begin{figure}[htb]
  \centering
  \includegraphics[width=0.49\linewidth]{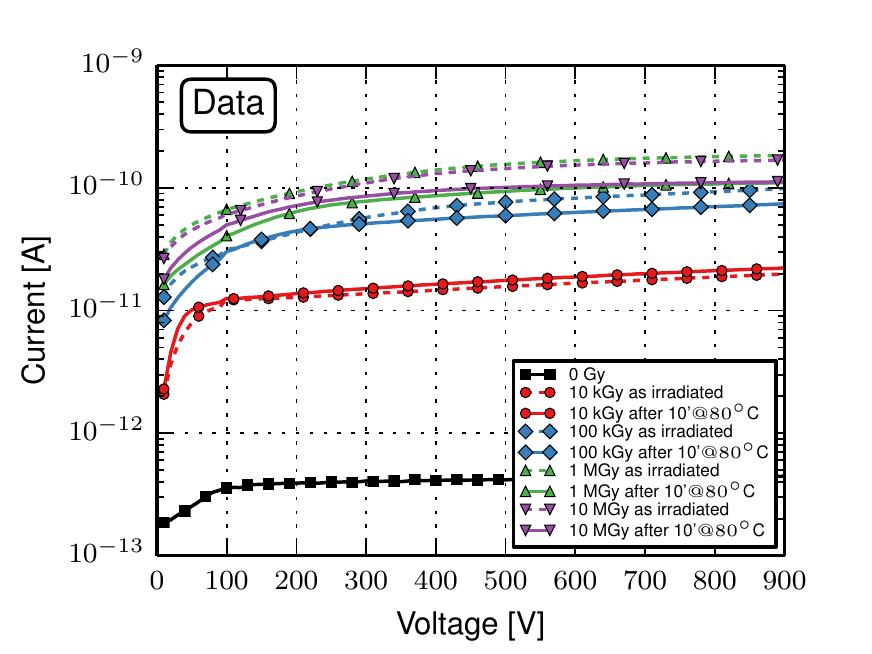}
    \includegraphics[width=0.49\linewidth]{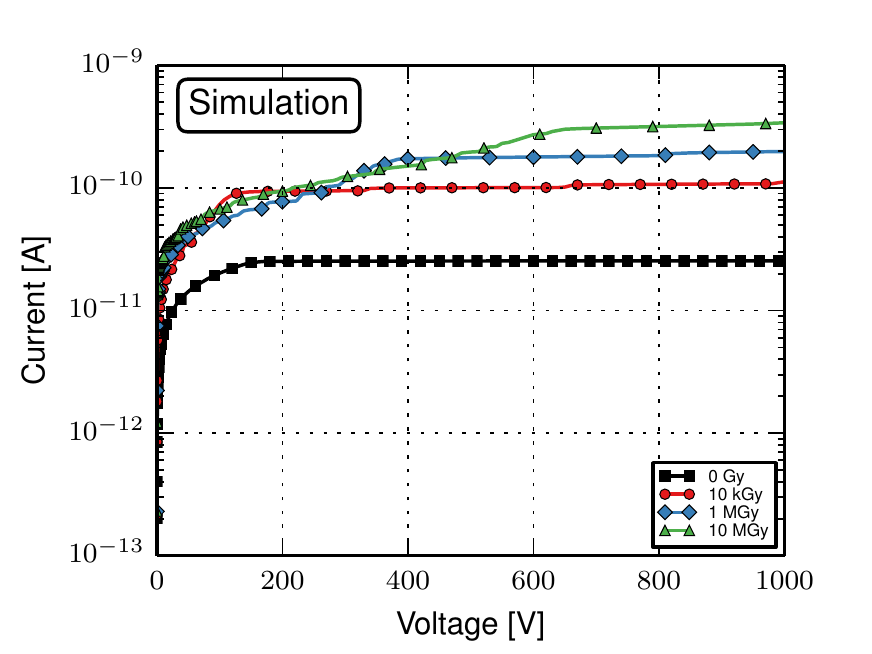}
  \caption{Top: Current per pixel of the 7$\times$7 test sensor for different dose values before and after annealing.
  Bottom: 3D simulations of the current per pixel for the different dose values with the values of $N_{ox}$ and $J_{surf}$ used for the optimization and given in table~\ref{tab:Radpar}.}
  \label{fig:Pixel_IV}
\end{figure}

 The top figure~\ref{fig:Pixel_IV} shows the currents per pixel measured at 20$^\circ$C in a dry atmosphere for the non-irradiated and the uniformly-irradiated sensor.
  No breakdown is observed up to a voltage of 900~V.
 After annealing for 10~minutes at 80$^\circ$C the maximal current per pixel is below 0.2~nA at 20$^\circ$C and below 3~pA at --20$^\circ$C.
  These values are well within the specifications shown in table~\ref{tab:Specs}.
 The bottom figure~\ref{fig:Pixel_IV} shows the predictions of the 3D simulations for comparison.
  Again, the values from the simulations and the measurements for the irradiated sensors are similar.
\begin{figure}[htb]
  \centering
  \includegraphics[width=0.6\linewidth]{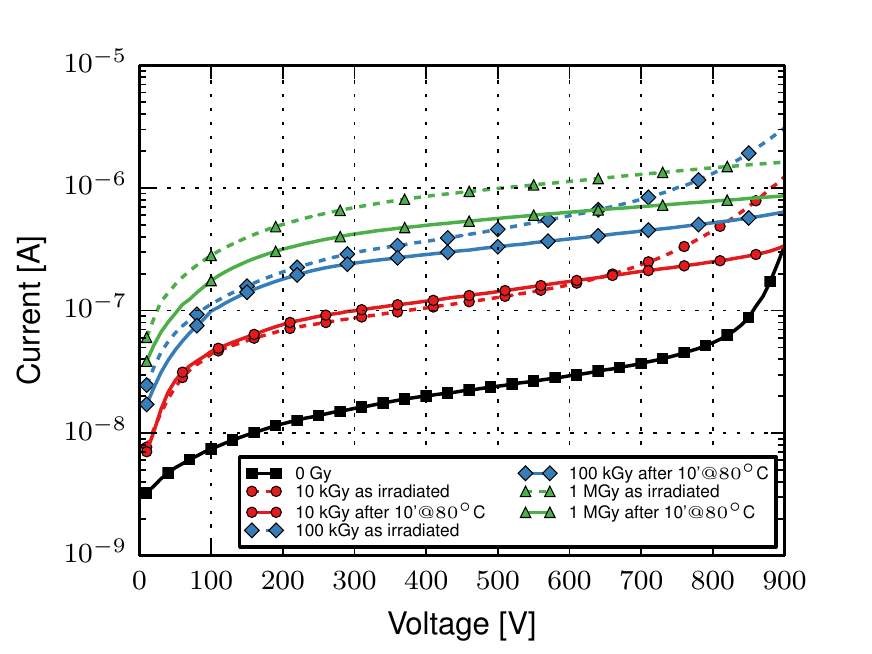}
  \caption{CCR current of the half-irradiated 7$\times$7 test sensor scaled to the full AGIPD sensor for
  different dose values before and after annealing.}
  \label{fig:CCR_half}
\end{figure}

\begin{figure}[htb]
  \centering
  \includegraphics[width=0.6\linewidth]{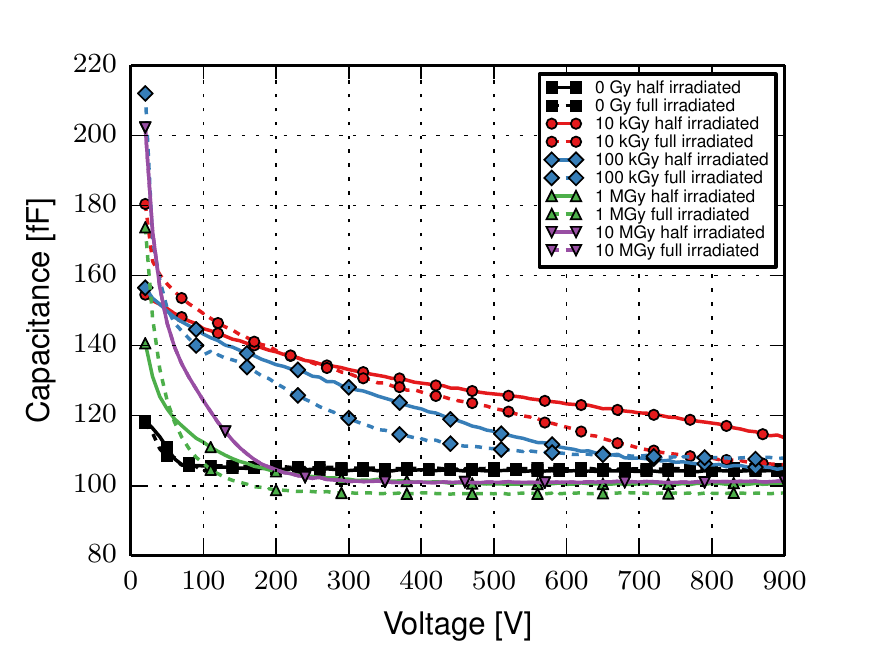}
  \caption{Inter-pixel capacitance $C_{int}$ at 1~MHz for the uniformly and half irradiated sensors after annealing for 10 minutes at 80$^\circ$C.}
  \label{fig:Interpixel}
\end{figure}

 To verify that the AGPID sensor also works if the irradiation is non-uniform, half of a sensor was irradiated up to a dose of 1~MGy whereas the other half was shielded by a 1~mm Ta absorber.
  The results for the CCR current for dose values up to 1~MGy, shown in figure~\ref{fig:CCR_half}, are as expected from the sum of the currents from a half-size irradiated and a half-size non-irradiated sensor.

 Finally, the inter-pixel capacitance $C_{int}$ as function of dose has been measured.
  It is found, that $C_{int}$ has some dependence on frequency, which however is less than 30~\% at 500~V, and will not be further discussed here.
 In figure~\ref{fig:Interpixel}, $C_{int}$ at 1~MHz after annealing for the uniformly and half-irradiated sensor, together with the results before irradiation, are presented.
  Before irradiation a value of 102~fF at 500~V is obtained, compared to the prediction of 98~fF.
 At a voltage of 500~V the maximum value of $C_{int}$ of 130~fF, reached at 10~kGy, is well below the specifications.

 We summarize: The AGIPD sensor fabricated by SINTEF and optimized for high operating voltage at high X-ray doses, achieves all specifications for the entire dose range for both uniform as well as for non-uniform irradiation.
  The reason for a soft breakdown for the non-irradiated sensor around 800~V is understood and does not present a problem.
 The measured current-voltage characteristics and the values for the inter-pixel capacitance as function of X-ray dose are quite similar to the predictions from the simulations.
  The breakdown voltage for the radiation-optimized design is significantly higher than for the standard design, and thus the optimization is considered a success.

\section{Conclusion}
 Imaging experiments at the European XFEL require silicon pixel sensors with unprecedented performance with respect to X-ray radiation tolerance at high operating voltages.
  The parameters which characterize the X-ray radiation damage of silicon sensors, which were obtained from test structures from four different vendors, have been implemented in TCAD simulations and both pixel and guard-ring layout optimized for radiation hardness.
 It is found that the optimization for X-ray radiation hardness is very different to the optimization for bulk radiation damage.
  Sensors according to the optimized parameters have been produced by SINTEF.
 Test structures and test sensors from the first production wafer have been irradiated up to an X-ray dose of 10~MGy.
  The measurements of the sensors before and after irradiation demonstrate, that the optimization has been successful, and that measured performance parameters like breakdown voltage, dark current and inter-pixel capacitance are close to the predictions by the simulations.

\section*{Acknowledgment}

 This work was performed within the AGIPD Project and was partially financed by the XFEL-Company.
  We would like to thank the AGIPD colleagues for the excellent collaboration.
 Support was also provided by the Helmholtz Alliance "Physics at the Terascale", and    J.~Zhang was supported by the Marie Curie Initial Training Network "MC-PAD".
  The authors also thank Thor-Erik Hansen and Niaz Ahmed from SINTEF for the pleasant and open collaboration and Alke Meents from DESY for setting up the irradiations in the P11 beam line of PETRA~III.



\begin{thebibliography}{10}

\bibitem{XFEL}
M.~Altarelli \emph{et~al.}, \emph{The European X-Ray Free-Electron Laser,
  Technical design report}.\hskip 1em plus 0.5em minus 0.4em\relax Hamburg,
  Germany: DESY, 2006.

\bibitem{Tschentscher:2011}
T.~Tschentscher \emph{et~al.}, ``Layout of the {X}-ray systems at the {European
  XFEL},'' European XFEL, Technical Note, 2011.

\bibitem{Graafsma:2009}
H.~Graafsma, ``Requirements for and development of 2 dimensional {X}-ray
  detectors for the european {X-ray Free Electron Laser} in {Hamburg},''
  \emph{J. Instrum.}, vol.~4, no.~12, p. P12011, 2009.

\bibitem{AGIPD}
{AGIPD}. [Online]. Available:
  \url{http://hasylab.desy.de/instrumentation/detectors/projects/agipd/index_eng.html}

\bibitem{Henrich:2011}
B.~Henrich \emph{et~al.}, ``The adaptive gain integrating pixel detector
  {AGIPD} a detector for the {European} {XFEL},'' \emph{Nucl. Instr. and Meth.
  A}, vol. 633, Supplement 1, pp. S11 -- S14, 2011.

\bibitem{Tove:1967}
P.~Tove and W.~Seibt, ``Plasma effects in semiconductor detectors,''
  \emph{Nucl. Instr. and Meth.}, vol.~51, no.~2, pp. 261 -- 269, 1967.

\bibitem{Becker:2010}
J.~Becker, D.~Eckstein, R.~Klanner, and G.~Steinbr{\"u}ck, ``Impact of plasma
  effects on the performance of silicon sensors at an {X}-ray {FEL},''
  \emph{Nucl. Instr. and Meth. A}, vol. 615, no.~2, pp. 230 -- 236, 2010.

\bibitem{Becker:Thesis}
J.~Becker, ``Signal development in silicon sensors used for radiation
  detection,'' Ph.D. dissertation, Universit{\"a}t Hamburg, Hamburg, 2010.

\bibitem{Zhang:2012}
J.~Zhang, I.~Pintilie, E.~Fretwurst, R.~Klanner, H.~Perrey, and J.~Schwandt,
  ``Study of radiation damage induced by 12~kev {X}-rays in {MOS} structures
  built on high-resistivity {\it n}-type silicon,'' \emph{J. Synchrotron Rad.},
  vol.~19, no.~3, pp. 340--346, 2012.

\bibitem{Zhang:2011}
J.~Zhang, E.~Fretwurst, R.~Klanner, H.~Perrey, I.~Pintilie, T.~Poehlsen, and
  J.~Schwandt, ``Study of {X}-ray radiation damage in silicon sensors,''
  \emph{J. Instrum.}, vol.~6, no.~11, p. C11013, 2011.

\bibitem{Zhang1:2012}
J.~Zhang, E.~Fretwurst, R.~Klanner, I.~Pintilie, J.~Schwandt, and M.~Turcato,
  ``Investigation of {X}-ray induced radiation damage at the {Si--SiO$_2$}
  interface of silicon sensors for the {European XFEL},'' \emph{J. Instrum.},
  vol.~7, no.~12, p. C12012, 2012.

\bibitem{Zhang:Thesis}
J.~Zhang, ``X-ray radiation damage studies and design of a silicon pixel sensor
  for science at the {XFEL},'' Ph.D. dissertation, Universit{\"a}t Hamburg,
  Hamburg, 2013.

\bibitem{Poehlsen:2012}
T.~Poehlsen, E.~Fretwurst, R.~Klanner, S.~Schuwalow, J.~Schwandt, and J.~Zhang,
  ``Charge losses in segmented silicon sensors at the {Si--SiO$_2$}
  interface,'' \emph{Nucl. Instr. and Meth. A}, vol. 700, pp. 22 -- 39, 2013.

\bibitem{Synopsis}
Synopsis {TCAD}. [Online]. Available: \url{http://www.synopsys.com/}

\bibitem{Schwandt:2012}
J.~Schwandt, E.~Fretwurst, R.~Klanner, I.~Pintilie, and J.~Zhang,
  ``Optimization of the radiation hardness of silicon pixel sensors for high
  {X}-ray doses using {TCAD} simulations,'' \emph{J. Instrum.}, vol.~7, no.~01,
  p. C01006, 2012.

\bibitem{Schwandt:2013}
------, ``Study of high-dose {X}-ray radiation damage of silicon sensors,'' in
  \emph{Proceedings of SPIE}, vol. 8777, 2013, p. 87770K.

\bibitem{Sintef}
{SINTEF ICT}. [Online]. Available: \url{http://www.sintef.no}

\bibitem{Poehlsen:2013}
T.~Poehlsen, J.~Becker, E.~Fretwurst, R.~Klanner, J.~Schwandt, and J.~Zhang,
  ``Study of the accumulation layer and charge losses at the {Si--SiO$_2$}
  interface in p$^+$n-silicon strip sensors,'' \emph{Nucl. Instr. and Meth. A},
  vol. 721, pp. 26 -- 34, 2013.

\bibitem{Poehlsen:Thesis}
T.~Poehlsen, ``Charge losses in silicon sensors and electric-field studies at
  the {Si--SiO$_2$} interface,'' Ph.D. dissertation, Universit{\"a}t Hamburg,
  Hamburg, 2013.

\bibitem{Bacchetta:2001}
N.~Bacchetta, D.~Bisello, A.~Candelori, M.~Rold, M.~Descovich, A.~Kaminski,
  A.~Messineo, F.~Rizzo, and G.~Verzellesi, ``Improvement in breakdown
  characteristics with multiguard structures in microstrip silicon detectors
  for {CMS},'' \emph{Nucl. Instr. and Meth. A}, vol. 461, pp. 204 -- 206, 2001.

\bibitem{Schwandt1:2012}
J.~Schwandt, E.~Fretwurst, R.~Klanner, and J.~Zhang, ``Design of the {AGIPD}
  sensor for the {European XFEL},'' \emph{J. Instrum.}, vol.~8, no.~01, p.
  C01015, 2013.

\bibitem{Grove:1967}
A.~Grove, \emph{Physics and Technology of Semiconductor Devices}.\hskip 1em
  plus 0.5em minus 0.4em\relax New York, {NY}: John Wiley \& Sons, 1967.

\end{thebibliography}
%

\end{document}